\documentclass[%
 reprint,
 prd,
superscriptaddress,
 amsmath,amssymb,
 aps,nofootinbib,
]{revtex4-1}

\usepackage{soul}
\usepackage{dsfont}
\usepackage{graphicx}
\usepackage{dcolumn}
\usepackage{bm}
\usepackage{color}
\usepackage{epsfig}
\usepackage{hyperref}
\usepackage{natbib}
\usepackage{float}
\usepackage{footmisc}
\usepackage{lipsum}
\usepackage{amsmath}
\usepackage{braket, xcolor}
\usepackage{hyperref}
\usepackage{comment}
\usepackage{MnSymbol}
\usepackage{enumitem}
\usepackage{longtable}
\usepackage{rotating,tabularx}
\hypersetup{colorlinks, citecolor=blue, linkcolor=blue, urlcolor=blue}

\newcommand{\be}{\begin{equation}}
\newcommand{\ee}{\end{equation}}
\newcommand{\beqa}{\begin{eqnarray}}
\newcommand{\eeqa}{\end{eqnarray}}

\begin{document}

\title{Measuring Pulsar Distances from Chirping Orbital Periods}

\author{Brady Egleston}
\email{beglesto@uw.edu}
\affiliation{Department of Physics, University of Maryland, College Park, Maryland 20742, USA}
\affiliation{Department of Physics, University of Washington, Seattle, WA 98195, USA}
\author{Reza~Ebadi}
\email{ebadi@umd.edu}
\affiliation{Department of Physics, University of Maryland, College Park, Maryland 20742, USA}
\affiliation{Department of Physics and Astronomy, The Johns Hopkins University, Baltimore, Maryland 21218, USA}
\affiliation{Department of Physics and Astronomy, University of Delaware, Newark, Delaware 19716, USA}
\author{Ronald Walsworth}
\email{walsworth@umd.edu}
\affiliation{Department of Physics, University of Maryland, College Park, Maryland 20742, USA}
\affiliation{Quantum Technology Center, University of Maryland, College Park, Maryland 20742, USA}
\affiliation{Department of Electrical and Computer Engineering, University of Maryland, College Park, Maryland 20742, USA\vspace{4pt}}


\begin{abstract}
The observed orbital period time derivative (or orbital ``chirp'') of a millisecond binary pulsar (MSP) encodes information about both the intrinsic properties of the binary system and its environment. Orbital chirp has contributions from intrinsic energy loss due to gravitational wave emission, kinematic effects due to motion in the plane of the sky, and dynamical effects due to galactic acceleration, with the latter two contributions depending on the MSP distance. We use orbital chirp data to infer distances to 21 MSPs; and for four of which we obtain smaller uncertainties than those reported in previous distance measurements. We incorporate multiple realistic galactic acceleration models to assess the sensitivity of the inferred distances to the choice of galactic gravitational potential, finding a significant dependence for four MSPs.
\end{abstract}

\maketitle

\section{Introduction \label{sec:intro}}

Timing of millisecond binary pulsars (MSPs) provides a unique avenue for precision astrophysics, with MSP frequency stability on par with human-made atomic clocks on long timescales (years) \cite{Hartnett_2011}. The applications span a wide range, including tests of gravitational theories \cite{Stairs:2003eg,Cornish:2017oic,Kramer:2021jcw,Meng:2025ezb,SKAPulsarScienceWorkingGroup:2025yqu}, studies of the quantum chromodynamics phase space \cite{Ozel:2016oaf,NANOGrav:2019jur}, galactic accelerometry \cite{Zakamska:2005hn,Phillips:2020xmf,Bovy_2020,Moran_2024}, and searches for long-period gravitational waves \cite{Detweiler:1979wn,NANOGrav:2023gor,Taylor:2025lxy} and dark matter \cite{Siegel:2007fz,Baghram:2011is,Porayko:2018sfa,EPTA:2023xxk,EuropeanPulsarTimingArray:2023egv,NANOGrav:2023hvm}. 

These studies inherently depend on accurate modeling and precise determinations of MSP parameters. In particular, distance to an MSP is both essential for many analyses and notoriously difficult to measure accurately \cite{Bell:1995ei,1990AJ....100..743F,Verbiest_2010,Smits_2011,Verbiest_2012}. A prominent example is the detection of gravitational waves from individual supermassive black hole binaries, which would benefit greatly from precise MSP distance measurements \cite{Zhu:2016zlx,Becsy:2025yzj,Kato:2023tfz}. 

A variety of methods have been developed to determine MSP distances. We describe some of these approaches in this section and highlight the associated challenges. See Table\,\ref{tab:nominal} for results from previous MSP distance analyses using the methods summarized here; and a comparison to the results from the present analysis.

The dispersion measure (DM) characterizes the wavelength-dependent delays in MSP pulse arrival times caused by dispersion in the interstellar medium at radio frequencies. The DM is defined as the column density of free electrons along the line of sight and therefore encodes the source distance through the upper limit of the line-of-sight integral \cite{YMW16}. Converting observed wavelength-dependent pulse delays into a distance estimate requires modeling the galactic electron density distribution \cite{TC93,Cordes:2002wz,Cordes:2003ik,YMW16}. However, this distribution is not well constrained, leading to typical uncertainties in the inferred MSP distances of about 20-40\%, and in some cases even larger \cite{Cordes:2002wz}.

Identification of a binary pulsar’s companion (e.g., a main sequence star or a white dwarf) in surveys such as Gaia enables another independent distance measurement through the companion’s observed parallax \cite{Jennings:2018clt,Moran_2023}. Optical imaging parallax is limited to sufficiently bright and nearby companions, such as main sequence stars \cite{van1984models}; and is challenging for many pulsar–white dwarf binaries at distances beyond kpc \cite{Moran_2023}. High-resolution radiofrequency imaging parallax of pulsars can also be achieved via Very Long Baseline Interferometry (VLBI) \cite{gwinn1986measurement,1995PhDT........20C,Fomalont:2004hr,Wang:2017lth}. This approach requires relatively long observing campaigns, and yields precise distance measurements for sufficiently bright and/or nearby pulsars \cite{Deller_2009,Smits_2011,Deller:2018zxz}. Current catalogs include VLBI-based distance measurements for several tens of pulsars \cite{Deller:2018zxz}.

Timing parallax (TP) relies on changes in the Earth-MSP distance as the Earth orbits the Sun, thereby inducing an annual modulation in MSP radio signal pulse arrival times \cite{Ryba_Taylor_1991,Shamohammadi:2024uvg}. While TP is present for all pulsars, an additional form of parallax information exists only for pulsars in binary systems: orbital parallax (OP). OP manifests as annual variations in the observed timing parameters, resulting from changes in the apparent orientation of the orbital plane due to the Earth’s motion around the Sun \cite{kopeikin1996proper,van2003annual}.

The observed MSP orbital period time derivative (or orbital ``chirp'') has also been used for distance determination \cite{Bell:1995ei} via the kinematic distance method. This approach exploits the apparent drift in orbital period caused by MSP proper motion \cite{Shklovskii_1970}. The effect is purely kinematic and can be understood as the time derivative of the observed MSP signal's Doppler shift (see Sec.\,\ref{sec:method}), distinct from intrinsic orbital decay driven by gravitational wave emission \cite{hulse1975discovery,Weisberg:2004hi}. In systems where the contribution from gravitational wave emission is well-understood or negligible, the kinematic distance method can, in principle, yield accurate distance estimates. Although environmental effects (such as tidal interactions and companion mass loss) may systematically bias this method, pulsars in binary systems with white dwarfs or neutron stars are less susceptible to these systematics due to their simpler environments \cite{Bell:1995ei,applegate1994orbital,Arzoumanian:1993qt,freire2012relativistic,Antoniadis:2013pzd}.

There is also a dynamical contribution to the observed orbital chirp from the relative line-of-sight acceleration between the MSP and the Earth \cite{shapiro1976galactic,Damour:1990wz}. This contribution is unavoidable, as the gradient of the galactic gravitational potential induces differential accelerations. Although less systematically accounted for in kinematic distance determinations, the spatial correlations between MSP accelerations have recently been analyzed to account for this dynamical contribution in the context of galactic accelerometry \cite{Zakamska:2005hn,Phillips:2020xmf,Bovy_2020,Moran_2024}.\footnote{See also Ref.~\cite{Ebadi:2024oaq} for galactic accelerometry using gravitational wave `clocks' (namely, inspiraling double white dwarfs) as analogs of MSP `clocks.'} A particular challenge in such galactic accelerometry studies is the limited precision of MSP distance measurements. Since MSPs sample the galactic potential at various locations, uncertainties in their positions make it more challenging to constrain global galactic parameters.

In this work, we focus on the reverse problem: we determine MSP distances from observed orbital chirps of MSP signals assuming realistic galactic gravitational potentials. Our method is similar to the kinematic distance technique, with the key difference that we systematically investigate the effect of the galactic acceleration contribution. In particular, we compute MSP distances using multiple acceleration models and compare the results, which allows us to assess the relative significance of the acceleration contribution to the orbital chirp. We also compare our results with existing MSP distance measurements reported in the literature. From this analysis, we gain valuable insight into the relative contributions of different sources of distance uncertainty, as well as the sensitivity of
the inferred distances to MSPs to each component of the orbital period derivative. The findings of this study may inform future efforts to improve the precision of galactic accelerometry measurements.

\section{Chirping orbital period \label{sec:method}}

The observed MSP orbital period $P_\textrm{b}^\textrm{\scriptsize obs}$ differs from its intrinsic value in the source frame $P_\textrm{b}$ due to the Doppler effect:
\begin{equation}
    P_\textrm{b}^\textrm{\scriptsize obs} = \bigg(1 + \frac{\Delta\mathbf{v} \cdot \hat{\mathbf{u}}}{c}\bigg)P_\textrm{b} +\mathcal{O}(\Delta v^2/c^2)
\end{equation}
where $\Delta \mathbf{v}$ is the relative velocity between the MSP and the Earth, $\hat{\mathbf{u}}$ is the unit vector along the line of sight, and $c$ is the speed of light. The time derivative of the orbital period thus receives contributions from three distinct terms:
\begin{equation}
    \dot{P}_\textrm{b}^\textrm{\scriptsize obs}= \dot{P}_\textrm{b}^\text{\scriptsize GW} + \dot{P}_\textrm{b}^\text{\scriptsize Shk} + \dot{P}_\textrm{b}^\text{\scriptsize Gal}
\end{equation}
where $\dot{P}_\textrm{b}^\text{\scriptsize GW} \propto \mathrm{d}P_\textrm{b}/\mathrm{d}t$, $\dot{P}_\textrm{b}^\text{\scriptsize Shk} \propto \mathrm{d}{\hat{\mathbf{u}}}/\mathrm{d}t$, and $\dot{P}_\textrm{b}^\text{\scriptsize Gal} \propto \mathrm{d}{\Delta \mathbf{v}}/\mathrm{d}t$. The first term corresponds to intrinsic energy loss due to gravitational wave emission \cite{Peters:1963ux,hulse1975discovery,Weisberg:2004hi}, the second term corresponds to the Shklovskii contribution \cite{Shklovskii_1970}, and the third term accounts for the effect of galactic acceleration \cite{shapiro1976galactic,Damour:1990wz}. Expressed in units of acceleration, this relation takes the form of
\begin{align}
    c\dot{P}_\textrm{b}^\textrm{\scriptsize obs}/P_\textrm{b} = c\dot{P}_\textrm{b}^\text{\scriptsize GW}/P_\textrm{b} + \mu^2D  + \Delta \mathbf{a}_{\rm gal} \cdot \hat{\mathbf{u}}
\end{align}
which we express as
\begin{equation}\label{eq:three_contributions}
    a^\textrm{obs} = a^\text{\scriptsize GW} + a^\text{\scriptsize Shk}(D) + a^\text{\scriptsize Gal}(D)
\end{equation}
where
\begin{align}
    a^\text{\scriptsize GW} &\equiv c\dot{P}_\textrm{b}^\text{\scriptsize GW}/P_\textrm{b}\\
    a^\text{\scriptsize Shk}(D) &\equiv \mu^2D \\
    a^\text{\scriptsize Gal}(D) &\equiv \Delta \mathbf{a}_{\rm gal}(D) \cdot \hat{\mathbf{u}}
\end{align}
Here, $\mu$ and $D$ are the MSP total proper motion and distance, respectively. The last two terms depend explicitly on MSP distance which we use to determine a best-fit value for $D$ from MSP observation data, as described below. 

We define the residual of the observed orbital period derivative data as:
\begin{equation}\label{eq:residual}
    \mathcal{R}(D) = a^\textrm{obs} - \Big[ a^\text{\scriptsize GW} + a^\text{\scriptsize Shk}(D) + a^\text{\scriptsize Gal}(D) \Big]\,.
\end{equation}
The goal of our analysis is to find a distance value that minimizes $\mathcal{R}^2(D)$.

\section{Galactic Acceleration Model\label{sec:gal_model}}
To compute $a^{\text{\scriptsize Gal}}(D)$, we use three different galactic gravitational potential models implemented in the \texttt{galpy} package \cite{Bovy_2015}:
\begin{itemize}
    \item \texttt{MWPotential2014}: adapted from \textcite{Bovy_2015}.
    \item \texttt{McMillan17}: adapted from \textcite{McMillan_2016}.
    \item \texttt{Cautun20}: adapted from \textcite{Cautun_2020}.
\end{itemize}
which we denote by B15, M17, and C20, respectively. We perform our analysis for all three models; however, unless explicitly noted otherwise, we adopt the C20 model as the default.

The B15 model is the simplest of the three, consisting of only three galactic components: a bulge, a thin stellar disk, and a dark matter halo. Its galactic gravitational potential is determined via a fit using phase-space data from individual stars~\cite{Bovy_2015}. The M17 model includes six components: a bulge, thin and thick stellar disks, two gas disks, and a dark matter halo. It makes use of kinematic data from galactic maser sources in its fitting procedure to determine the galactic potential~\cite{McMillan_2016}. The C20 model has seven components: a bulge, thin and thick stellar disks, two gas disks, a circumgalactic medium, and a dark matter halo. It is optimized by fitting to the Milky Way rotation curve, thereby yielding the galactic potential. We note that C20 employs a contracted dark matter halo, in contrast to the Navarro-Frenk-White (NFW) halo profiles used in B15 and M17~\cite{Cautun_2020}.

In addition to having fewer components than the other models, B15 has a smaller bulge in both radius and mass, and the absence of a thick disk makes the resulting galactic gravitational potential more spherical and less oblate. In contrast, M17 and C20 are largely consistent within uncertainties for their shared components, although minor differences exist. In C20, the stellar disks are slightly wider but less dense, the dark matter halo has a slightly larger scale radius, and the bulge is marginally more massive, while the total stellar mass is slightly lower compared to M17.

\begin{figure*}[t]
    \centering
    \includegraphics[width=\linewidth]{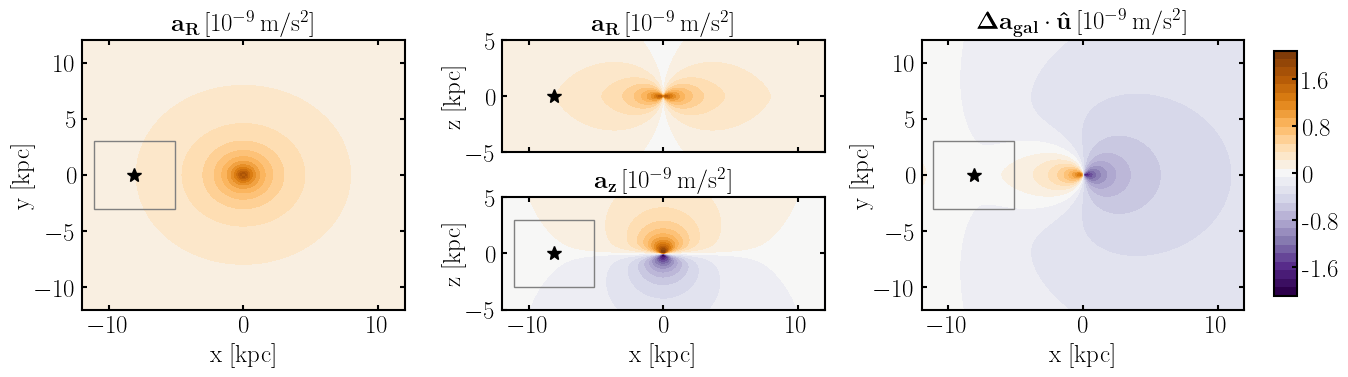}
    \includegraphics[width=\linewidth]{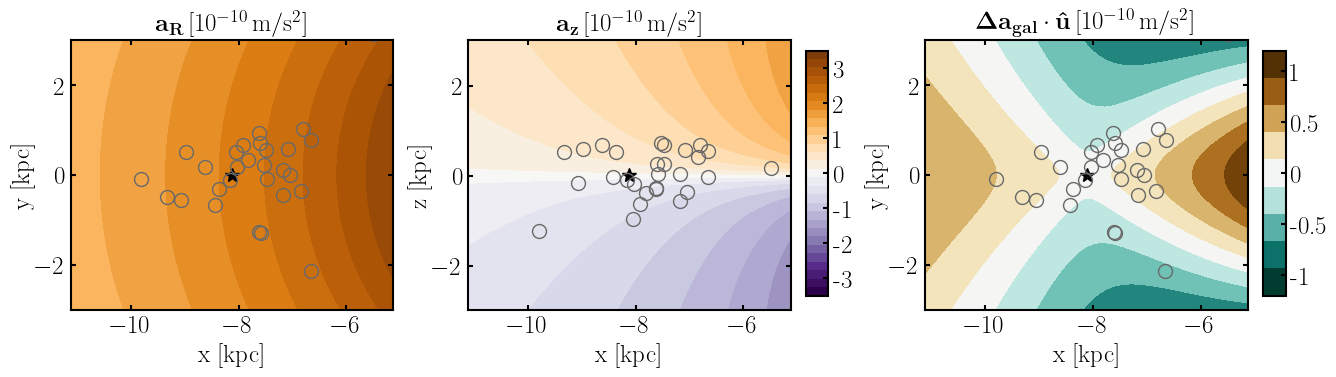}
    \caption{\textbf{Top:} Galactic acceleration profile determined from the C20 galactic gravitational potential model. {\it Left}: Radial acceleration in the x-y plane (galactic plane). {\it Middle–Top}: Radial acceleration in the x-z plane. {\it Middle–Bottom}: Vertical acceleration in the x-z plane. {\it Right}: Line-of-sight differential acceleration. Spatial variations in acceleration are on the order of $10^{-9}\,{\rm m/s^2}$. The location of the Sun is marked with a star ($\star$), and the gray squares outline the region highlighted in the zoomed-in profiles below. The middle panels demonstrate that radial acceleration is dominant over vertical acceleration ($a_R>a_z$) in the solar neighborhood.
    \textbf{Bottom:} Zoomed-in view of the galactic acceleration profile in the solar neighborhood. Local variations in acceleration are at the level of $10^{-10}\,{\rm m/s^2}$.
    We note that the radial acceleration is dominant over the vertical acceleration ($a_R>a_z$) in this region of the galaxy. The color scale in the right plot is selected to approximately reflect the sensitivity of pulsar orbital period measurements to acceleration effects. Open circles mark the positions of the millisecond binary pulsars (MSPs) analyzed in this work.}
    \label{fig:galaxy}
\end{figure*}

Fig.\,\ref{fig:galaxy} visualizes the C20 galactic potential both on the large galactic scale (top panels) and locally within the solar neighborhood (bottom panels). The variation of the galactic acceleration across the galaxy is on the order of $10^{-9}\,{\rm m/s^2}$, while locally, in the solar neighborhood, it is smaller, about $10^{-10}\,{\rm m/s^2}$. We also indicate the locations of the MSPs in our dataset (see Sec.\,\ref{sec:pulsars}), most of which lie within a few kpc of the Sun. From the bottom-right panel of Fig.\,\ref{fig:galaxy}, it is evident that a robust reconstruction of the local acceleration map requires an orbital chirp precision of better than $0.3\times10^{-10}\,{\rm m/s^2}$, after correcting for gravitational wave and Shklovskii effects with a similar or better level of precision.

\begin{figure*}[t]
    \centering
    \includegraphics[width=\linewidth]{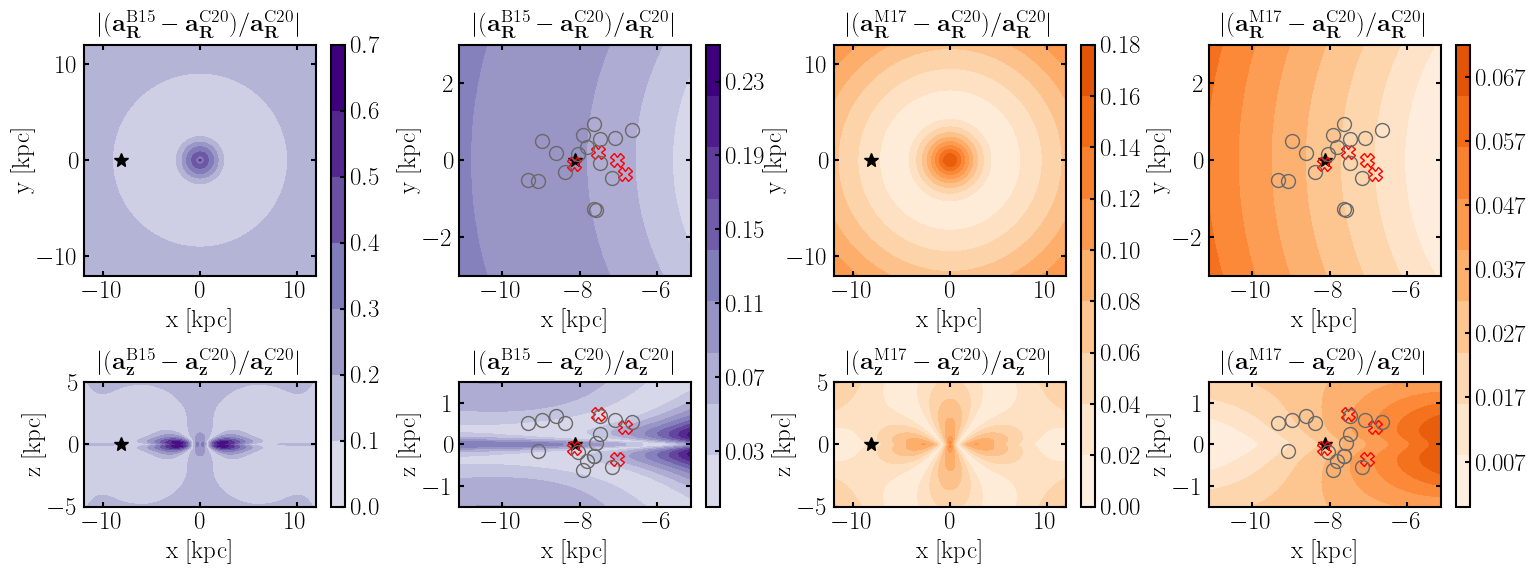}
    \caption{Fractional differences in predicted gravitational accelerations between galactic potential models. The maps show the relative differences of the M17 and B15 models compared to C20 as the reference model. Generically, the fractional difference is $\lesssim 10\%$. Open circles indicate the positions of millisecond binary pulsars (MSPs) in our dataset. Red crosses indicate MSPs B1534+12, J0437–4715, J1600–3053, and J1909–3744, for which we find that the inferred distances are significantly sensitive to the choice of galactic gravitational potential model.}
    \label{fig:gal_model_comparison}
\end{figure*}

Fig.\,\ref{fig:gal_model_comparison} compares the three galactic potential models considered here: B15, M17, and C20. The maximum fractional variation of galactic acceleration for B15 relative to C20 is approximately 0.7 on galactic scales and 0.2 in the solar neighborhood. For M17 compared to C20, these variations are smaller, around 0.15 and 0.05, respectively. Despite these relatively modest differences in galactic acceleration magnitude, the models exhibit distinct spatial patterns, which are critical for distinguishing between them using MSP data. However, this distinction is degenerate with the precision of MSP distance measurements. We also highlight four pulsars (B1534+12, J0437-4715, J1600-3053, and J1909-3744) in this figure for which the distance posterior distributions show significant sensitivity to the choice of galactic potential; see Fig.\,\ref{fig:gal_sensitivity}. Interestingly, although the acceleration in the M17 model is generally closer to that of the C20 model on overall scales, the predicted line-of-sight acceleration in the B15 model is closer to the C20 model at the sky locations of three of these four pulsars (with J0437–4715 being the exception), as determined by comparing the values on the right-hand side of Eq.~(\ref{eq:three_contributions}).

\section{Pulsar Dataset 
\label{sec:pulsars}}

In earlier work, some of us utilized a catalog of 13 millisecond binary pulsars (MSPs) for galactic accelerometry \cite{Phillips:2020xmf}. Since then, other groups have expanded on this effort by employing larger MSP samples to investigate slow drifts in parameters such as the orbital period, both for galactic accelerometry (see, e.g., Ref.\,\cite{Moran_2024}) and for pico-Hz gravitational wave detection \cite{DeRocco:2023qae,Zheng:2025tcm}. In this work, we adopt the extended catalog presented in Ref.\,\cite{Moran_2024}, using 26 of the 29 MSPs in that catalog, plus two pulsars from our original catalog not in \cite{Moran_2024}: J1756-2251 and J1829+2456. Three pulsars from \cite{Moran_2024} are excluded (J1455–3330, J1603–7207, and J2129–5721) because pulsar or companion masses for these systems could not be found in the literature. We also adjust several elements of the dataset from \cite{Moran_2024}, including corrections to the uncertainties on proper motion and updates to incorporate more recent PTA measurements for some MSPs. The resulting updated MSP dataset analyzed here is shown in Table~\ref{tbl:binary_psr}. The data for all 31 MSPs including the three that are excluded from our analysis can be found in the file {\tt Pulsar\_Params\_Expanded.csv} attached to this manuscript.

This sample is considerably smaller than the full catalog of known MSPs.  This is because obtaining a complete orbital solution with sufficient precision is not feasible for most MSPs in the full catalog. For the MSPs included in Table\,\ref{tbl:binary_psr}, we can compute all three contributions to the orbital chirp in Eq.\,\eqref{eq:three_contributions}.

In Fig.\,\ref{fig:a_contributions}, we illustrate the interplay among the three acceleration contributions ($a^\text{\scriptsize GW}$, $a^\text{\scriptsize Shk}$, and $a^\text{\scriptsize Gal}$) for each MSP in our catalog. The relative amplitude of each contribution compared to the observed chirp (top panels) indicates its importance in the measurement, and hence the parameters associated with that contribution may be constrained with higher precision. Similarly, comparing the magnitude of a given contribution to the measurement uncertainty (bottom panels) indicates whether it lies above the detection threshold; that is, whether the corresponding parameters are detectable at all in our analysis.

We distinguish between MSPs with convergent distance estimates and those without in Fig.\,\ref{fig:a_contributions} (see the next section). A general trend emerges: MSPs exhibiting a significant Shklovskii contribution tend to have more precisely determined distances in this work, provided that this contribution exceeds the total measurement uncertainty. We next discuss in detail the results and their interpretations.

\section{Results and Discussion \label{sec:results}}

\begin{figure*}[t]
    \centering
    \includegraphics[width=\linewidth]{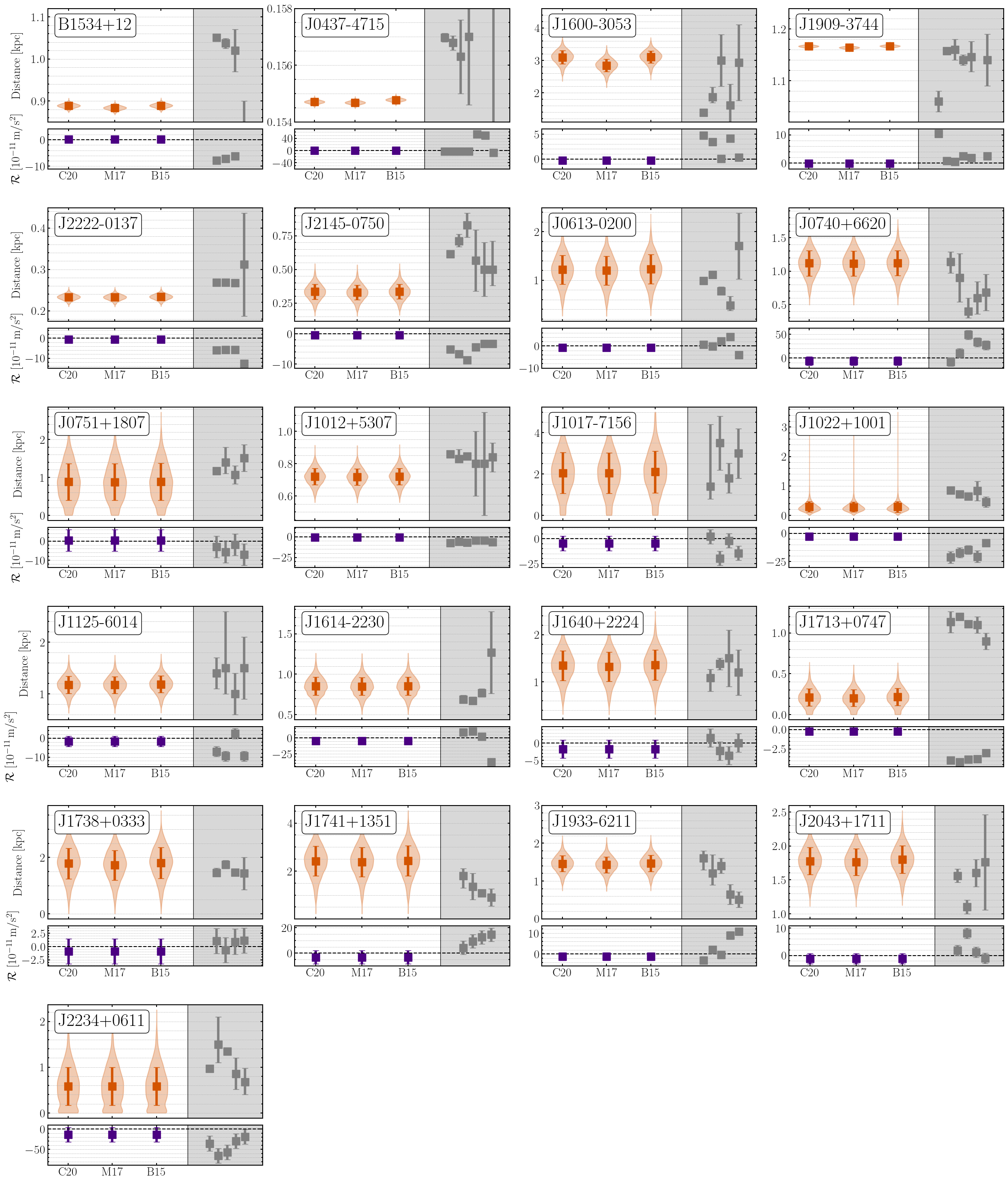}
    \caption{Inferred distances for the millisecond binary pulsars (MSPs) analyzed in this work. The top panel for each MSP shows distance measurement results derived using the C20, M17, and B15 galactic potential models. Violin plots represent the distribution of distances obtained by minimizing the orbital chirp residual. In the gray shaded area we show distance measurements reported in the literature (see Table\,\ref{tab:nominal} for references). The bottom panel shows the residual of the orbital chirp in Eq.\,\eqref{eq:residual} for each MSP distance. Our minimization procedure yields improved agreement between the model predictions and the observed orbital chirps.}
    \label{fig:results}
\end{figure*}

For the MSP dataset in Table~\ref{tbl:binary_psr}, we optimize the orbital chirp residual equation (Eq.\,\eqref{eq:residual}) to obtain best-fit distance values. Our analysis employs Monte Carlo sampling of the six-dimensional MSP parameter space, followed by minimization of the residual function using the \texttt{minimize\_scalar} routine in the \texttt{scipy} package \cite{2020SciPy-NMeth}.\footnote{We also performed a maximum likelihood analysis using Markov chain Monte Carlo (MCMC) methods and found results nearly identical to those obtained with the simpler residual minimization approach. We therefore present the results of the latter analysis.} This procedure yields posterior distributions for MSP distances and residuals, shown in Figures\,\ref{fig:results} and \ref{fig:gal_sensitivity}. We repeat the analysis for three different galactic acceleration models (B15, M17, and C20).

We present the results of our MSP distance determinations obtained via residual minimization in Table\,\ref{tab:nominal}. This table also includes previously reported distances in the literature. Our distance determinations produce improved uncertainties for four MSPs: J0437-4715, J1909-3744, J1017-7156, and J1022+1001. We also produce new distance determinations with comparable uncertainties to previous efforts for a further three MSPs: B1534+12, J0740+6620, and J1933-6211.

The results are further visualized in Fig.\,\ref{fig:results}, which also shows the residuals as defined in Eq.\,\eqref{eq:residual}. We find that our minimization method provides an improved solution to Eq.\,\eqref{eq:three_contributions}, as indicated by the smaller residuals compared to those associated with the previously-reported MSP distances. Finally, in Fig.\,\ref{fig:gal_sensitivity}, we present MSP distance distributions from our analysis. This figure illustrates the sensitivity of each MSP’s inferred distance to the choice of galactic potential model.

\begin{figure*}[t]
    \centering
    \includegraphics[width=\linewidth]{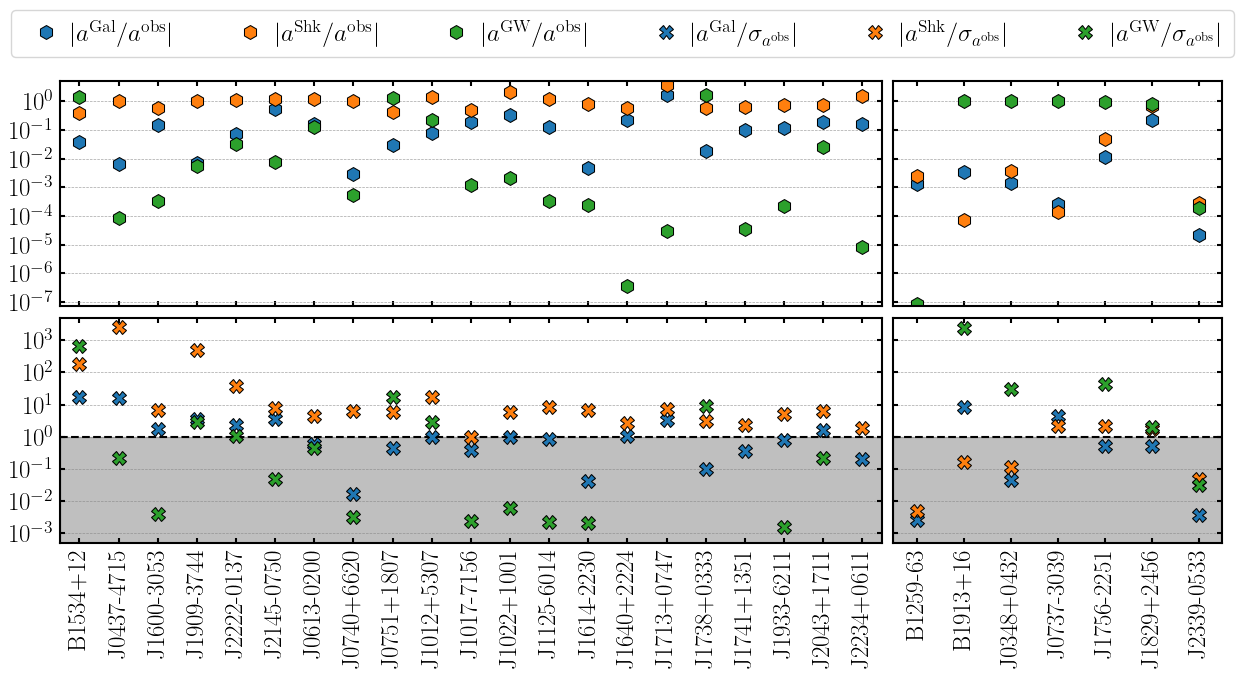}
    \caption{Comparison of the three terms ($a^{\rm GW}$, $a^{\rm Shk}$, and $a^{\rm Gal}$) contributing to the observed apparent acceleration $a^{\rm obs}$ for millisecond binary pulsars (MSPs) analyzed in this work. \textbf{Top:} Relative contribution of each term to the total observed acceleration, shown as the ratios $|a^{\rm GW}/a^{\rm obs}|$, $|a^{\rm Shk}/a^{\rm obs}|$, and $|a^{\rm Gal}/a^{\rm obs}|$. \textbf{Bottom:} Amplitude of each term relative to the measurement uncertainty in the observed acceleration $|a^{\rm GW}/\sigma_{a^{\rm obs}}|$, $|a^{\rm Shk}/\sigma_{a^{\rm obs}}|$, and $|a^{\rm Gal}/\sigma_{a^{\rm obs}}|$. The dashed line marks the level at which any term's contribution equals the measurement uncertainty; below this line (shaded in gray) approximately indicates contributions that are not statistically significant for our analysis. \textbf{Left:} MSPs for which we obtain convergent distance measurements. For all of these, the Shklovskii term contributes significantly and lies above the significance threshold in the bottom panel. While the galactic acceleration term is subdominant for many MSPs, it still exceeds the uncertainty level in several cases and thus is significant in the analysis (Sec.\,\ref{sec:results}). \textbf{Right:} MSPs for which distance measurements do not converge. These are dominantly binary systems for which the approximation $a^{\rm obs} \approx a^{\rm GW}$ holds well. As a result, our analysis lacks sensitivity to distance for these sources. There are, however, two notable exceptions: PSR~J1829 and PSR~J2339, where distance-dependent terms are significant; however, the large uncertainty in the intrinsic gravitational wave contribution (indicated by the green cross near the dashed line) limits our ability to constrain the distance.}
    \label{fig:a_contributions}
\end{figure*}

\begin{figure*}[t]
    \centering
    \includegraphics[width=\linewidth]{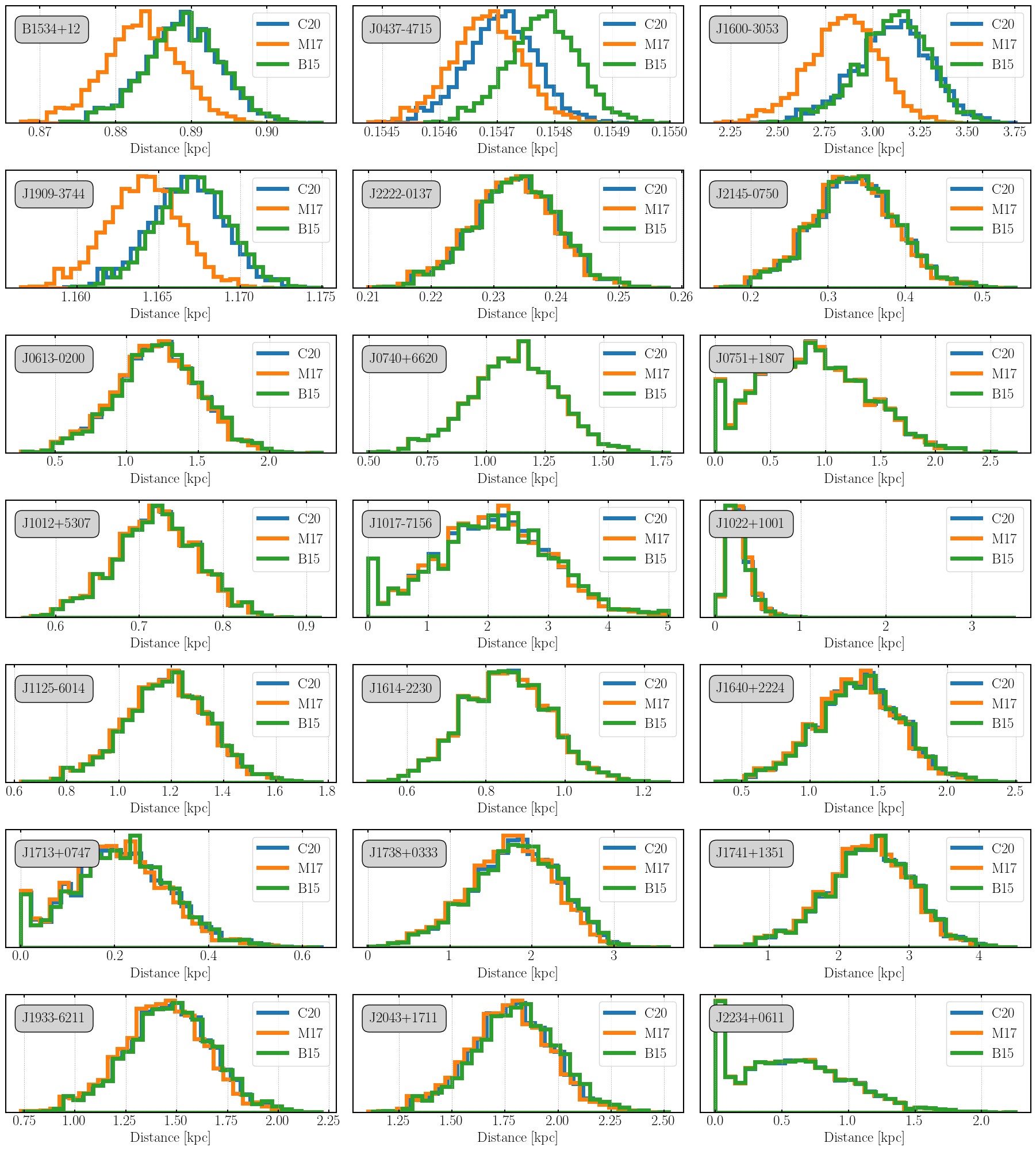}
    \caption{Normalized probability distribution functions of millisecond binary pulsar (MSP) distances that minimize the residual in Eq.\,\eqref{eq:residual}. B1534+12 and J0437, with the highest $a^{\rm Gal}/\sigma_{a^{\rm obs}}\sim 10$ (Fig.\,\ref{fig:a_contributions}), are sensitive to the galactic gravitational potential. Among MSPs with $a^{\rm Gal}/\sigma_{a^{\rm obs}}\sim1$, only J1600 and J1909 show significant sensitivity. The remaining MSPs with $a^{\rm Gal}/\sigma_a^{\rm obs}<1$ have minimal sensitivity to the galactic gravitational potential.}
    \label{fig:gal_sensitivity}
\end{figure*}

There are four MSPs with the highest precision for the orbital period chirp, characterized by $a^{\rm obs}/\sigma_{a^{\rm obs}} > 100$ in Fig.\,\ref{fig:a_contributions}: B1534+12 (also known as J1537+1155), J0437–4715, J1909-3744, and B1913+16 (also known as J1915+1606, the well-known Hulse-Taylor binary). As shown in Fig.\,\ref{fig:results}, J0437–4715, B1534+12, and J1909-3744 yield the highest significance level for distance determinations using our method. The Hulse-Taylor binary, however, does not produce a convergent distance estimate. This is because its intrinsic orbital decay due to gravitational wave emission (which is independent of distance) dominates the observed orbital chirp. In hindsight, this result is consistent with the role this system played in providing the first indirect confirmation of gravitational wave emission from stellar binaries \cite{hulse1975discovery}.

Two of the four MSPs mentioned above as having highest precision for the orbital chirp, B1534+12 and J0437-4715, also exhibit significant sensitivity to the choice of galactic potential model (Fig.\,\ref{fig:gal_sensitivity}). This result is expected since our analysis finds that the galactic contribution to their acceleration is substantial compared to the measurement uncertainty of the observed pulsar signal over time, with $a^{\rm Gal}/\sigma_{a^{\rm obs}}\gtrsim 10$ (Fig.\,\ref{fig:a_contributions}). For J1909–3744, the galactic acceleration contribution is less than the measurement uncertainty.

The next best-measured orbital period chirps correspond to J2222-0137, J0751+1807, J1012+5307, J0438+0432, and J1756–2251, with $10 < a^{\rm obs}/\sigma_{a^{\rm obs}} < 100$ for each of these MSPs (Fig.\,\ref{fig:a_contributions}). Our analysis for J1012+5307 and J2222-0137 yields high-significance distance measurements (Fig.\,\ref{fig:results}); for J0751+1807 we find a low-significance distance measurement since the gravitational wave contribution is dominant; and for J0438+0432 and J1756–2251 our analysis does not produce convergent distances, for the same reason as for the Hulse–Taylor binary. 

For J1713+0747, the galactic acceleration and Shklovskii terms are both larger than the observed orbital chirp and have opposite signs; therefore, the inferred distance relies on a precise cancellation between these terms. Because this cancellation must be fine-tuned, the inferred distance in our analysis is biased and has a tail at zero distance in the posterior distribution. Similar behavior found for J1022+1001, J0751+1807, and J2234+0611 can be understood in the same way. The remaining MSP results can be interpreted by examining the comparison of contributions shown in Fig.~\ref{fig:a_contributions}, as discussed in this section.

\section{Conclusion}

In this work, we analyze orbital chirp data for a dataset of millisecond binary pulsars (MSPs) having complete orbital solutions; and obtain convergent distance estimates for 21 MSPs. We further examine the three primary contributions to the observed orbital chirp (gravitational wave emission, the Shklovskii effect, and galactic acceleration); and identify the dominant contribution for each system. We assess potential explanations for why some MSPs yield convergent results while others do not, which can inform follow-up studies of additional MSPs. Furthermore, we systematically study the impact of galactic acceleration on binary pulsar orbital chirp; and find four MSPs for which distance estimates have significant sensitivity  to the choice of galactic potential. These results can inform future studies using MSPs for precision galactic accelerometry. 

Before closing, we note a key distinction between parallax-based MSP distances and those derived from orbital chirp: the parallax timing signal has a constant amplitude, while acceleration-like timing signals ($a^\text{\scriptsize Shk}$ and $a^\text{\scriptsize Gal}$) grow as $\propto t^2$. As a result, the improvement in signal-to-noise ratio with additional timing data scales as $\propto \sqrt{t}$ for parallax-based distances, but as $\propto t^{2.5}$ for the method presented here, consistent with the scaling in the kinematic distance method \cite{Bell:1995ei}.

\section*{Acknowledgments}
We thank David F. Phillips and Aakash Ravi for insightful discussions during the early stages of this project, which contributed to the conception of the main idea. We also thank Jo Bovy and Aakash Ravi for a careful read of the manuscript and comments. 
This work is supported by, or in part by, the Department of Energy under Grant. No. DESC0021654; the U.S. Army Research Laboratory under Contract No. W911NF2420143; and the University of Maryland Quantum Technology Center.
R.E. is supported by the John Templeton Foundation Award No. 63595 and the University of Delaware Research Foundation NSF Grant No. PHY-2515007. The work of R.E. was also supported by the Grant 63034 from the John Templeton Foundation. R.E. gratefully acknowledges the Pacific Postdoctoral Program at the Dark Universe Science Center, University of Washington, where part of this work was carried out. The Pacific Postdoctoral Program is supported by a grant from the Simons Foundation (SFI-MPS-T-Institutes-00012000, ML).

\onecolumngrid
\begingroup
\renewcommand{\arraystretch}{1.3}
\setlength{\tabcolsep}{1pt}
\begin{center}
\begin{sidewaystable}
    \caption{Summary of distances determinations from literature for the 21 millisecond binary pulsars (MSPs) for which we find convergent results using our analysis. The final column reports our result using the C20 galactic gravitational potential model. We use the following abbreviations for MSP distance determination methods: orbital parallax (OP), orbital parallax with kinematic term only (KN), timing parallax (TP), bootstrap (BS), dispersion measure (DM), spectroscopy (SP), and VLBI.}
    \begin{tabular}{c c c c c c c c c c c}
        \hline \hline
        Pulsar Name & \multicolumn{6}{c}{Distance [kpc], Method, Reference} & This Work [kpc] \\
        \hline
        B1534+12 & 1.051(5) OP \cite{Fonseca:2014qla} & 1.037(12) OP \cite{Fonseca:2012thd} & 1.02(5) OP \cite{Stairs:2002muf} & 0.7(2) DM \cite{Stairs:2002muf} & & & 0.889(5)\\
        J0437-4715 & 0.15696(11) OP \cite{Reardon:2024tfk} & 0.15679(25) OP \cite{Reardon:2015kba} & 0.1563(13) VLBI \cite{Deller:2008awv} & 0.1570(24) KN \cite{Verbiest:2008nbr} & 0.121$_{-0.008}^{+0.013}$ TP \cite{Jennings:2018clt} & 0.16 DM \cite{Jennings:2018clt} &  0.15471(6)\\
        J1600-3053 & 1.39(4) OP \cite{EuropeanPulsarTimingArray:2023sqx} & 1.87$_{-0.18}^{+0.30}$ TP \cite{Reardon:2021pph} & 3.0(8) KN \cite{Reardon:2021pph} & 1.62 DM \cite{Ord:2006bge} & 2.93 DM \cite{Ord:2006bge} & & 3.1(2)\\
        J1909-3744 & 1.06(2) TP \cite{EuropeanPulsarTimingArray:2023sqx} & 1.158(3) OP \cite{Liu:2020yvr} & 1.16(2) TP \cite{Liu:2020yvr} & 1.14(1) TP \cite{Perera:2019sca} & 0.460 DM \cite{Hotan:2006fjp} & 1.14(5) TP \cite{Hotan:2006fjp} & 1.167(2)\\
        J2222-0137 & 0.2686$_{-0.0009}^{+0.0010}$ VLBI \cite{Ding:2024isc} & 0.2680(12) VLBI \cite{Guo:2021oug} & 0.2673$_{-0.0009}^{+0.0012}$ VLBI \cite{Cognard:2017xyr} & 0.312 DM \cite{Cognard:2017xyr} & & & 0.233(7)\\
        J2145-0750 & 0.62$_{-0.02}^{+0.00}$ VLBI \cite{Deller:2018zxz} & 0.71$_{-0.04}^{+0.05}$ TP \cite{Reardon:2021pph} & 0.83(9) KN \cite{Reardon:2021pph} & 0.566 DM \cite{Reardon:2015kba} & 0.5 DM \cite{Reardon:2015kba} & 0.50$_{-0.12}^{+0.21}$ TP \cite{Lohmer:2004xeo} & 0.33(6)\\
        J0613-0200 & 0.99(5) TP \cite{EuropeanPulsarTimingArray:2023sqx} & 1.11(5) TP \cite{Perera:2019sca} & 0.777$_{-0.070}^{+0.084}$ TP \cite{Desvignes:2016yex} & 0.480$_{-0.110}^{+0.190}$ TP \cite{Hotan:2006fjp} & 1.705 DM \cite{Hotan:2006fjp} & & 1.2(3)\\
        J0740+6620 & 1.14$_{-0.15}^{+0.17}$ OP \cite{Fonseca:2021znv} & 0.9 DM \cite{Fonseca:2021znv} & 0.4$_{-0.1}^{+0.2}$ TP \cite{NANOGrav:2018ouk} & 0.6 DM \cite{Fonseca:2021znv} & 0.68 DM \cite{Guillemot:2016vgo} & & 1.12(19)\\
        J0751+1807 & 1.17$_{-0.05}^{+0.06}$ TP \cite{EuropeanPulsarTimingArray:2023sqx} & 1.4$_{-0.3}^{+0.4}$ TP \cite{Perera:2019sca} & 1.07(24) TP \cite{Desvignes:2016yex} & 1.51(35) TP \cite{Guillemot:2016vgo} & & & 0.9(5)\\
        J1012+5307 & 0.86(2) VLBI \cite{Ding:2020qla} & 0.83$_{-0.02}^{+0.06}$ BS \cite{Ding:2020qla} & 0.845(14) TP+GAIA \cite{Moran_2023} & 0.8(2) TP \cite{lazaridis2009} & 0.80 DM \cite{Jennings:2018clt} & 0.84(9) SP \cite{Callanan:1998dvr} & 0.72(5)\\
        J1017-7156 & 1.4$_{-0.6}^{+3.0}$ TP \cite{Reardon:2021pph} & 3.5(1.3) KN \cite{Reardon:2021pph} & 1.8 DM \cite{Price:2021bud} & 3.0 DM \cite{Price:2021bud} & & & 2(1)\\
        J1022+1001 & 0.85$_{-0.05}^{+0.06}$ TP \cite{EuropeanPulsarTimingArray:2023sqx} & 0.72$_{-0.02}^{+0.01}$ VLBI \cite{Deller:2018zxz} & 0.64$_{-0.08}^{+0.06}$ TP \cite{Reardon:2021pph} & 0.83 DM \cite{Deller:2018zxz} & 0.45 DM \cite{Deller:2018zxz} & & 0.29$_{-0.13}^{+0.12}$\\
        J1125-6014 & 1.4(3) KN \cite{Reardon:2021pph} & 1.5$_{-0.5}^{+1.1}$ \cite{Reardon:2021pph} & 1.0 DM \cite{Smith:2019cdc} & 1.5 DM \cite{Smith:2019cdc} & & & 1.17(16)\\
        J1614-2230 & 0.69$_{-0.04}^{+0.05}$ TP \cite{Perera:2019sca} & 0.67$_{-0.04}^{+0.05}$ TP \cite{NANOGrav:2018ouk} & 0.77(5) TP \cite{Guillemot:2016vgo} & 1.27 DM \cite{NANOGrav:2016bay} & & & 0.85(11)\\
        J1640+2224 & 1.08$_{-0.19}^{+0.28}$ TP \cite{EuropeanPulsarTimingArray:2023sqx} & 1.39$_{-0.13}^{+0.11}$ VLBI+TP \cite{Ding:2023fli} & 1.5 DM \cite{Perera:2019sca} & 1.2 DM \cite{Ding:2023fli} & & & 1.35(32)\\
        J1713-0747 & 1.136(13) TP \cite{EuropeanPulsarTimingArray:2023sqx} & 1.20(3) TP \cite{Perera:2019sca} & 1.108$_{-0.033}^{+0.035}$ TP \cite{Desvignes:2016yex} & 1.1(1) TP \cite{Splaver:2005aqx} & 0.9(1) DM \cite{Espinosa:2013lnu} & & 0.21(11)\\
        J1738+0333 & 1.45$_{-0.07}^{+0.17}$ DM \cite{Koljonen:2024egi} & 1.74$_{-0.13}^{+0.15}$ VLBI+TP \cite{Ding:2023fli} & 1.47(10) TP \cite{freire2012relativistic} & 1.43 DM \cite{Desvignes:2016yex} & & & 1.78(54)\\
        J1741+1351 & 1.8$_{-0.3}^{+0.5}$ TP \cite{NANOGrav:2018ouk} & 1.36 DM \cite{Kirichenko:2020gct} & 1.08(5) TP \cite{Espinosa:2013lnu} & 0.9 DM \cite{Kirichenko:2020gct} & & & 2.4(6)\\
        J1933-6211 & 1.6$_{-0.2}^{+0.3}$ OP+KN \cite{Geyer:2023dhy} & 1.2$_{-0.3}^{+0.5}$ OP \cite{Geyer:2023dhy} & 1.4(2) KN \cite{Geyer:2023dhy} & 0.65 DM \cite{Geyer:2023dhy} & 0.51 DM \cite{Geyer:2023dhy} & & 1.46(21)\\
        J2043+1711 & 1.56(10) TP \cite{Donlon:2025bvt} & 1.1(1) TP \cite{Perera:2019sca} & 1.6(2) TP \cite{NANOGrav:2018ouk} & 1.76 DM \cite{Guillemot:2016vgo} & & & 1.8(2)\\
        J2234+0611 & 0.97(4) TP \cite{Stovall:2019fru} & 1.5$_{-0.4}^{+0.6}$ TP \cite{NANOGrav:2018ouk} & 1.35(5) TP \cite{Antoniadis:2016nfy} & 0.86 DM \cite{Stovall:2019fru} & 0.68 DM \cite{Stovall:2019fru} & & 0.58$_{-0.48}^{+0.43}$\\
    \end{tabular}
    \label{tab:nominal}
\end{sidewaystable}
\end{center}
\endgroup

\clearpage
\setlength{\tabcolsep}{-1.7pt}
\begin{center}
\begin{longtable}{ l  c c c c c c c c c c c c c c }
\caption{Millisecond binary pulsar (MSP) dataset considered in this work. \label{tbl:binary_psr}}\\ \\ \hline\hline
Pulsar name & B1259-63 \cite{MillerJones:2018gdv,Shannon:2014pnh} & B1534+12 \cite{Fonseca:2014qla} & B1913+16 \cite{Weisberg2016,Weisberg2008} & J0348+0432 \cite{Saffer:2025ezw, Antoniadis:2013pzd}  \\ \hline
Galactic longitude, $l$ [deg] & 304.183553 & 19.847526 & 49.967651 & 183.336841\\
Galactic latitude, $b$ [deg] & -0.991583 & 48.341431 & 2.121891 & -36.773627\\
Orbital period, $P_{\text{b}}$ [days] & 1236.724525(6) & 0.420737298881(2) & 0.322997448918(3) & 0.102424061371(9)\\
Derivative of $P_{\text{b}}$, $\dot{P}_{\text{b}}$ [10$^{-12}$ s s$^{-1}$] & 14000(7000) & -0.1366(3) & -2.423(1) & -0.217(7)\\
Pulsar mass, $M_1$ $[M_\odot]$ & 2(1) & 1.3330(2) & 1.438(1) & 1.806(37)\\
Companion mass, $M_2$ $[M_\odot]$ & 23(8) & 1.3455(2) & 1.390(1) & 0.154(3)\\
Orbital eccentricity, $e$ $[10^{-6}]$ & 869879.70(6) & 273677.52(7) & 617134.0(4) & 4.7(10)\\
Total proper motion, $\mu$ [mas year$^{-1}$] & 7.03(3) & 25.328(12) & 0.77(11) & 4.270(92)\\
Companion &High Mass Star&Neutron Star&Neutron Star&White Dwarf\\
\hline\hline
Pulsar name & J0437-4715 \cite{Reardon:2024tfk, Perera:2019sca} & J0613-0200 \cite{Perera:2019sca,EuropeanPulsarTimingArray:2023sqx} & J0737-3039A/B \cite{Kramer:2021jcw} & J0740+6620 \cite{Fonseca:2021znv, NANOGrav:2018ouk}  \\ \hline
Galactic longitude, $l$ [deg] & 253.394523 & 210.413052 & 245.2357 & 149.729731\\
Galactic latitude, $b$ [deg] & -41.963302 & -9.304907 & -4.5049 & 29.599367\\
Orbital period, $P_{\text{b}}$ [days] & 5.74104582(16) & 1.198512575217(10) & 0.1022515592973(1) & 4.76694461936(8)\\
Derivative of $P_{\text{b}}$, $\dot{P}_{\text{b}}$ [10$^{-12}$ s s$^{-1}$] & 3.7329(15) & 0.026(7) & -1.247920(78) & 1.2(2)\\
Pulsar mass, $M_1$ $[M_\odot]$ & 1.418(44) & 1.75(55) & 1.338185(14) & 2.08(7)\\
Companion mass, $M_2$ $[M_\odot]$ & 0.221(4) & 0.14(3) & 1.248868(13) & 0.2530(55)\\
Orbital eccentricity, $e$ $[10^{-6}]$ & 19.1819(11) & 4.50(9) & 87777.023(61) & 5.990(29)\\
Total proper motion, $\mu$ [mas year$^{-1}$] & 140.913(1) & 10.51(1) & 3.305(33) & 32.666(200)\\
Companion &White Dwarf&White Dwarf&Radio Pulsar&White Dwarf\\
\hline\hline
Pulsar name & J0751+1807 \cite{Perera:2019sca,Desvignes:2016yex} & J1012+5307 \cite{Perera:2019sca,Ding:2020qla,Matasanchez:2020cyr} & J1017-7156 \cite{Reardon:2021pph} & J1022+1001 \cite{Perera:2019sca, Reardon:2021pph}  \\ \hline
Galactic longitude, $l$ [deg] & 202.729676 & 160.347147 & 291.558116 & 231.794541\\
Galactic latitude, $b$ [deg] & 21.08584 & 50.857836 & -12.55327 & 51.100642\\
Orbital period, $P_{\text{b}}$ [days] & 0.263144270792(7) & 0.604672723085(3) & 6.5119041(9) & 7.8051353(6)\\
Derivative of $P_{\text{b}}$, $\dot{P}_{\text{b}}$ [10$^{-12}$ s s$^{-1}$] & -0.0350(25) & 0.052(4) & 0.4(2) & 0.21(7)\\
Pulsar mass, $M_1$ $[M_\odot]$ & 1.64(15) & 1.72(16) & 2.0(8) & 1.44(44)\\
Companion mass, $M_2$ $[M_\odot]$ & 0.16(1) & 0.165(15) & 0.27(40) & 0.52(24)\\
Orbital eccentricity, $e$ $[10^{-6}]$ & 3.3(5) & 1.1(1) & 142.4(6) & 97.02(4)\\
Total proper motion, $\mu$ [mas year$^{-1}$] & 13.675(294) & 25.521(60) & 10.105(12) & 19.647(3696)\\
Companion &White Dwarf&White Dwarf&White Dwarf&White Dwarf\\
\hline\hline
Pulsar name & J1125-6014 \cite{Reardon:2021pph} & J1600-3053 \cite{EuropeanPulsarTimingArray:2023sqx,Reardon:2021pph} & J1614-2230 \cite{Perera:2019sca,NANOGrav:2018ouk,NANOGrav:2023dke} & J1640+2224 \cite{EuropeanPulsarTimingArray:2023sqx,NANOGrav:2023dke}  \\ \hline
Galactic longitude, $l$ [deg] & 292.504194 & 344.090438 & 352.635721 & 41.051005\\
Galactic latitude, $b$ [deg] & 0.89368 & 16.451381 & 20.192109 & 38.271186\\
Orbital period, $P_{\text{b}}$ [days] & 8.7526036468(4) & 14.3484635(2) & 8.68661955647(8) & 175.460664578(9)\\
Derivative of $P_{\text{b}}$, $\dot{P}_{\text{b}}$ [10$^{-12}$ s s$^{-1}$] & 0.7(1) & 0.36(3) & 1.7(2) & 9(2)\\
Pulsar mass, $M_1$ $[M_\odot]$ & 1.5(2) & 2.060(425) & 1.937(14) & 2.80(175)\\
Companion mass, $M_2$ $[M_\odot]$ & 0.31(3) & 0.29(2) & 0.494(2) & 0.45(18)\\
Orbital eccentricity, $e$ $[10^{-6}]$ & 0.615(11) & 173.728(2) & 1.330(7) & 797.277(3)\\
Total proper motion, $\mu$ [mas year$^{-1}$] & 17.126(14) & 6.984(11) & 33.196(994) & 11.526(7)\\
Companion &White Dwarf&White Dwarf&White Dwarf&White Dwarf\\
\hline\hline
Pulsar name & J1713+0747 \cite{Reardon:2021pph} & J1738+0333 \cite{Ding:2023fli,freire2012relativistic} & J1741+1351 \cite{Perera:2019sca,NANOGrav:2018ouk,NANOGrav:2023dke} & J1756-2251 \cite{Ferdman:2014rna}  \\ \hline
Galactic longitude, $l$ [deg] & 28.750557 & 27.721282 & 37.885185 & 6.498658\\
Galactic latitude, $b$ [deg] & 25.222836 & 17.742237 & 21.640797 & 0.94801\\
Orbital period, $P_{\text{b}}$ [days] & 67.825139(3) & 0.3547907398724(13) & 16.3353478289(2) & 0.31963390143(3)\\
Derivative of $P_{\text{b}}$, $\dot{P}_{\text{b}}$ [10$^{-12}$ s s$^{-1}$] & 0.2(1) & -0.0170(31) & 1.330(365) & -0.229(5)\\
Pulsar mass, $M_1$ $[M_\odot]$ & 1.28(8) & 1.460(55) & 1.14(34) & 1.341(7)\\
Companion mass, $M_2$ $[M_\odot]$ & 0.283(9) & 0.1810(75) & 0.220(45) & 1.230(7)\\
Orbital eccentricity, $e$ $[10^{-6}]$ & 74.9408(5) & 0.34(11) & 10.00(1) & 180569.4(2)\\
Total proper motion, $\mu$ [mas year$^{-1}$] & 6.293(2) & 8.675(8) & 11.649(20) & 2.42$\pm 20$ \footnote{The proper motion in declination has only an upper bound $\mu_\delta < 20$ mas/yr; we assume $\mu_\delta = 0 \pm 20$ mas/yr.}\\
Companion &White Dwarf&White Dwarf&White Dwarf&Neutron Star\\
\pagebreak
\hline\hline
Pulsar name & J1829+2456 \cite{Haniewicz:2021jro} & J1909-3744 \cite{Perera:2019sca,EuropeanPulsarTimingArray:2023sqx,Reardon:2021pph} & J1933-6211 \cite{Geyer:2023dhy} & J2043+1711 \cite{Perera:2019sca,NANOGrav:2023dke,NANOGrav:2016bay}  \\ \hline
Galactic longitude, $l$ [deg] & 53.3426 & 359.730783 & 334.4309 & 61.918822\\
Galactic latitude, $b$ [deg] & 15.6119 & -19.59581 & -28.6315 & -15.31289\\
Orbital period, $P_{\text{b}}$ [days] & 1.176027952868(11) & 1.533449475874(1) & 12.819406716(1) & 1.4822908095(1)\\
Derivative of $P_{\text{b}}$, $\dot{P}_{\text{b}}$ [10$^{-12}$ s s$^{-1}$] & -0.029(12) & 0.509(1) & 0.7(1) & 0.102(12)\\
Pulsar mass, $M_1$ $[M_\odot]$ & 1.306(4) & 1.486(11) & 1.40(25) & 1.62(10)\\
Companion mass, $M_2$ $[M_\odot]$ & 1.299(4) & 0.2081(9) & 0.43(5) & 0.173(10)\\
Orbital eccentricity, $e$ $[10^{-6}]$ & 139143.74(13) & 0.104(6) & 1.26(2) & 4.87(10)\\
Total proper motion, $\mu$ [mas year$^{-1}$] & 9.52(7) & 37.020(6) & 12.42(3) & 12.257(18)\\
Companion &Neutron Star&White Dwarf&White Dwarf&White Dwarf\\
\hline\hline
Pulsar name & J2145-0750 \cite{Perera:2019sca,Reardon:2021pph,Deller:2016srt} & J2222-0137 \cite{Guo:2021oug} & J2234+0611 \cite{Stovall:2019fru} & J2339-0533 \cite{Pletsch:2015tbk,Romani:2011lvf}  \\ \hline
Galactic longitude, $l$ [deg] & 47.776711 & 62.018455 & 72.99 & 81.348942\\
Galactic latitude, $b$ [deg] & -42.083628 & -46.075288 & -43.01 & -62.476158\\
Orbital period, $P_{\text{b}}$ [days] & 6.83890261495(9) & 2.445759995469(6) & 32.00140163(8) & 0.1930984018(30)\\
Derivative of $P_{\text{b}}$, $\dot{P}_{\text{b}}$ [10$^{-12}$ s s$^{-1}$] & 0.13(2) & 0.2510(76) & 3.1(25) & -166(1)\\
Pulsar mass, $M_1$ $[M_\odot]$ & 1.8(4) & 1.81(3) & 1.3530(155) & 1.40(4)\\
Companion mass, $M_2$ $[M_\odot]$ & 0.83(6) & 1.312(9) & 0.2980(135) & 0.075(7)\\
Orbital eccentricity, $e$ $[10^{-6}]$ & 19.34(3) & 380.92(1) & 129274.035(8) & 210.0(1)\\
Total proper motion, $\mu$ [mas year$^{-1}$] & 13.148(51) & 45.061(41) & 27.10(2) & 31.016(9457)\\
Companion &White Dwarf&White Dwarf&White Dwarf&Low Mass Star\\
\end{longtable}
\end{center}

\clearpage

\twocolumngrid
\bibliography{references}

\clearpage
\newpage

\end{document}